\documentclass[preprints,article,accept,moreauthors,pdftex]{Definitions/mdpi} 

\firstpage{1} 
\pubvolume{1}
\issuenum{1}
\articlenumber{0}
\pubyear{2021}
\copyrightyear{2020}
\datereceived{} 
\dateaccepted{} 
\datepublished{} 
\hreflink{https://doi.org/} 

\Title{The ionization energies of dust-forming metal oxide clusters }

\TitleCitation{Metal oxide cluster ionization nergies}

\Author{David Gobrecht $^{1,}$*\orcidA{}, Jan Philip Sindel $^{1,2,3}$, Helena Lecoq-Molinos $^{1}$ and Leen Decin $^{1}$}

\AuthorNames{David Gobrecht, Jan Philip Sindel, Helena Lecoq-Molinos and Leen Decin}

\AuthorCitation{Gobrecht, D.; Sindel, J.-P.; Lecoq-Molinos, H.; Decin, L.}

\address{%
$^{1}$ \quad Institute of Astronomy, KU Leuven, Celestijnenlaan 200D, 3001 Leuven, Belgium; dave@gobrecht.ch\\
$^{2}$ \quad SUPA,  School of Physics \& Astronomy, University of St Andrews, North Haugh, St Andrews, KY169SS, UK\\
$^{3}$ \quad Centre for Exoplanet Science, University of St Andrews, North Haugh, St Andrews, KY169SS, UK}

\corres{Correspondence: dave@gobrecht.ch}

\abstract{Stellar dust grains are predominantly composed of mineralic, anorganic material forming in the circumstellar envelopes of oxygen-rich AGB stars. 
However, the initial stage of the dust synthesis, or its nucleation, is not well understood.  
In particular, the chemical nature of the nucleating species, represented by molecular clusters, is uncertain.\
We investigate the vertical and adiabatic ionization energies of four different metal-oxide clusters by means of density functional theory. They include clusters of magnesia (MgO)\textsubscript{n}, silicon monoxide (SiO)\textsubscript{n}, alumina (Al\textsubscript{2}O\textsubscript{3})\textsubscript{n}, and titania (TiO\textsubscript{2})\textsubscript{n} with stoichiometric sizes of n=1$-$8.\
The magnesia, alumina and titania clusters show relatively little variation in their ionization energies with respect to the cluster size n, ranging from 7.1$-$8.2 eV for (MgO)\textsubscript{n}, 8.9$-$10.0 eV for (Al\textsubscript{2}O\textsubscript{3})\textsubscript{n}, and 9.3$-$10.5 eV  for (TiO\textsubscript{2})\textsubscript{n}. In contrast, the (SiO)\textsubscript{n} ionization energies decrease with size n, starting from 11.5 eV for n=1, and decreasing to 6.6 eV for n=8.\
Therefore, we set constraints on the stability limit for neutral metal-oxide clusters to persist
ionization through radiation or high temperatures and for the nucleation to proceed via neutral-neutral reactions.}


\begin{document}


\section{Introduction}
The formation of dust in AGB (Asymptotic Giant Branch) star envelopes is not well understood \citep{2018A&ARv..26....1H}.
In particular, the dust nucleation around oxygen-rich AGB stars lacks of knowledge, since it is not clear, which chemical species initiate and partake in the condensation process 
\citep{1995A&AS..112..143H}. 
The composition of a dust grain, once it has formed and grown to a size on the (sub-)micron scale,  
is known from the analysis of meteoritic stardust \citep{2003TrGeo...1...17Z}. 
A great variety of different materials are identified in presolar stardust grains, with silicon-oxide compounds, or silicates, being the major constituent \citep{2013LPI....44.2367N,Hoppe_2015}. 
The silicates are found to be magnesium-rich (rather than iron-rich) with a mixed olivine-pyroxene stoichiometry of which only a fraction is crystalline \citep{1998A&A...339..904J,2007A&A...462..667M,2001A&A...369..132K,doi:10.1111/j.1365-2966.2011.20255.x,doi:10.1021/acsearthspacechem.9b00139}.\\  
The condensation temperatures of silicate material are known to range around 900$-$1300 K around AGB stars\citep{doi:10.1021/je800560b,2013A&A...555A.119G}. 
However, large circumstellar dust grains with sizes of 0.6 $\mu$m form within two stellar radii, where the temperatures are higher \citep{2012Natur.484..220N}. Moreover, a dust formation zone close to the star is required to drive the mass loss of an AGB star \citep{2003A&A...399..589H}.
Therefore, it is thus likely that silicates do not nucleate on their own (i.e. {\it homogeneously}), but {\it heterogeneously} on the surface of previously formed seed nuclei. The 
nature (composition, size, crystallinity) of these seed nuclei is still a matter of debate 
\citep{2019MNRAS.tmp.2040B}.
Owing to their minor contribution to the total dust mass, seed nuclei can also be made of less abundant elements. However, a crucial requirement for seed nuclei is to be refractory. 
Alumina and titania fulfill this criterium and in addition, they are promising carriers of a spectral dust feature, commonly observed at a wavelength of 13 $\mu$m in oxygen-rich AGB stars \citep{1999A&A...352..609P,2003ApJS..149..437P,2003ApJ...594..483S}. 
Albeit silicon monoxide and magnesia are less refractory and hence less likely to trigger dust nucleation in AGB stars, they represent basic building blocks of silicates and are included in our study.
Previous studies have addressed the four nucleation candidates MgO, SiO, Al\textsubscript{2}O\textsubscript{3}, and TiO\textsubscript{2}, respectively, and described the properties of the related neutral clusters in astrophysical environments theoretically \citep{1997A&A...320..553K,bhatt2005,C6CP03629E,refId0D,2015A&A...575A..11L}, as well as experimentally \citep{demyk04,C2CP23432G,doi:10.1063/1.4894406}.
We note, however, that it is probable that chemically-heterogeneous compounds involving 
several clustering species could play a role in the rich gas-phase mixture of AGB circumstellar envelopes.\\
Nucleation is commonly presumed to proceed via neutral-neutral reactions.
An aspect, that is often ignored in the studies of dust nucleation, is the degree of ionization, or, the fraction of matter that is charged.
AGB stars are rather cool (T=2000$-$3000 K) resulting in a stellar radiation field with a relatively mild UV contribution. However, pulsation-induced shocks 
propagate periodically through the AGB envelopes enhancing the temperatures locally and temporarily \citep{1985ApJ...299..167B}. In addition, some AGB stars reside in binary systems with a hot companion star providing a source of ionizing UV radiation, as for example Mira B. Moreover, interstellar UV radiation can have an ionizing effect on the gas, particularly at larger distances to the star \citep{1983A&A...128..212M}. 
Therefore,  we aim at constraining the conditions under which oxygen-rich dust can nucleate through neutral metal-oxide clusters and at which temperatures their related cations become important.\\
This study represents a continuation of \citep{2020MNRAS.tmp.2040B} who investigated the nucleation of neutral clusters of four metal oxide families (MgO, SiO, Al\textsubscript{2}O\textsubscript{3}, TiO\textsubscript{2}). Here, we address the corresponding ionization energies and their related cations allowing us to assess whether a fast ion-molecule chemistry can take place and compete with a comparatively slow neutral-neutral nucleation.

This paper is organized as follows.
In Section \ref{methods}, we describe the methods to calculate the ionization potentials. 
Section \ref{results} addresses the results obtained for the four metal oxide clusters.
Finally, we summarize and discuss our findings in Section \ref{discus}.

 
\section{Methods}
\label{methods}
Our study focuses on four different metal oxide nucleation candidates that were presented in the study of \citep{2019MNRAS.tmp.2040B}. 
They include clusters of (MgO)\textsubscript{n}, (SiO)\textsubscript{n}, (Al\textsubscript{2}O\textsubscript{3})\textsubscript{n}, and (TiO\textsubscript{2})\textsubscript{n} for sizes n=1$-$8.
We perform Density Functional Theory (DFT) calculations of the neutral cluster and its singly ionized cation to compute the ionization potential, or ionization energy $E_{i}(X_n)$, for each species $X$ and each size $n$:\\
\begin{equation}
E_{i}(X_{n}) = E(X_{n}^{+}) - E(X_{n})
\end{equation}

where E(X$_{n}^{+}$) and E(X$_{n}$) are the potential energies of X$_{n}^{+}$ and X$_{n}$, respectively.
For reasons of comparability all DFT calculations were performed on the same level of theory. We choose the B3LYP/6-311+(d) hybrid density functional / basis set \citep{1993JChPh..98.1372B}, as it provides a reasonable accuracy at an affordable computational cost, and is suitable for inorganic metal oxides \citep{doi:10.1080/00268970500179651}. All DFT calculations were performed with the software suite of Gaussian 09 \citep{g09}.\\
For our calculations, we assume that the neutral metal oxide clusters are present in the form of their global minimum (GM) candidate structure.
The corresponding geometries originate from \citep{doi:10.1021/jp412820z} for MgO, from \citep{C6CP03629E} for SiO, from \citep{LI2012125,GOBRECHT2018138} for Al\textsubscript{2}O\textsubscript{3}, and from \citep{C6NR05788H} for TiO\textsubscript{2}, and are subsequently optimized. The optimization is performed for both, the neutral cluster $X$ and its cation $X$\textsuperscript{+}, and includes a vibration frequency analysis. 
This analysis allows us to discriminate between true minima with only real frequencies and transition states (TS) with an imaginary vibration mode.\\ 
Apart from the energy and the charge, also the spin multiplicity of the cations is different from their neutral counterparts. Here, all cations are presumed to be in a doublet
state (i.e. with a spin multiplicity of 2), since the neutral metal oxide clusters are singlet states, apart from the triplet Al\textsubscript{2}O\textsubscript{3} monomer. Test calculations on small sized (n=1$-$4) metal oxide cations in quartet states (i.e. with a spin multiplicity of 4) show higher potential energies than their corresponding doublet states (if they converge at all).
We distinguish between {\it vertical} and {\it adiabatic} ionization energies.
The vertical ionization energies are evaluated according to single point energies. In contrast, adiabatic ionization energies take into account structural rearrangements that are caused by the ionization process (i.e. by removing an electron) and are consequently lower than the vertical ionization energies.\\ 
\section{Results}
\label{results}
\subsection{Vertical ionization}
In Table \ref{tab1} the vertical (and adiabatic) ionization energies of the four oxide families (magnesia, silicon monoxide, alumina, titania) are listed as a function of cluster size n.

\begin{specialtable}[H] 
\caption{The vertical and adiabatic ionization energies, E\textsuperscript{v} and E\textsuperscript{a}, given in units of eV for the four considered metal oxide clusters MgO, SiO, (Al\textsubscript{2}O\textsubscript{3}), (TiO\textsubscript{2}) as a function of cluster size n=1$-$8\label{tab1}.}
\begin{tabular}{r | cc | cc | cc | cc}
\toprule
n &  \multicolumn{2}{c}{\textbf{(MgO)\textsubscript{n}}} & \multicolumn{2}{c}{\textbf{(SiO)\textsubscript{n}}} & \multicolumn{2}{c}{\textbf{(Al\textsubscript{2}O\textsubscript{3})\textsubscript{n}}} & \multicolumn{2}{c}{\textbf{(TiO\textsubscript{2})\textsubscript{n}}} \\
  & E\textsuperscript{v} &  E\textsuperscript{a} & E\textsuperscript{v} & E\textsuperscript{a} & E\textsuperscript{v} & E\textsuperscript{a} & E\textsuperscript{v} & E\textsuperscript{a} \\
\midrule
1 & 7.86 & 7.75 & 11.49   &  11.49 & 9.39   & 9.15 & 9.81   & 9.67  \\
2 & 7.82 & 7.56 & 9.22    &   9.21 & 9.81   & 9.52 & 10.50  & 10.23 \\
3 & 8.19 & 8.18 & 9.01    &   8.74 & 9.98   & 9.45 & 9.92   & 9.77  \\
4 & 7.90 & 7.37 & 8.39    &   8.34 & 9.88   & 9.52 & 10.54  & 10.43 \\
5 & 7.61 & 7.09 & 8.20    &   7.66 & 9.74   & 9.48 & 10.25  & 9.32 \\
6 & 7.94 & 7.63 & 7.89    &   7.30 & 9.73   & 9.54 & 10.32  & 9.34 \\
7 & 7.74 & 7.23 & 8.05    &   7.35 & 9.72   & 9.48 & 9.40   & 9.27 \\
8 & 7.54 & 7.16 & 7.04    &   6.62 & 9.13   & 8.90 & 10.08  & 9.04  \\
\bottomrule
\end{tabular}
\end{specialtable}

\begin{figure}[H]
\includegraphics[width=10.0 cm]{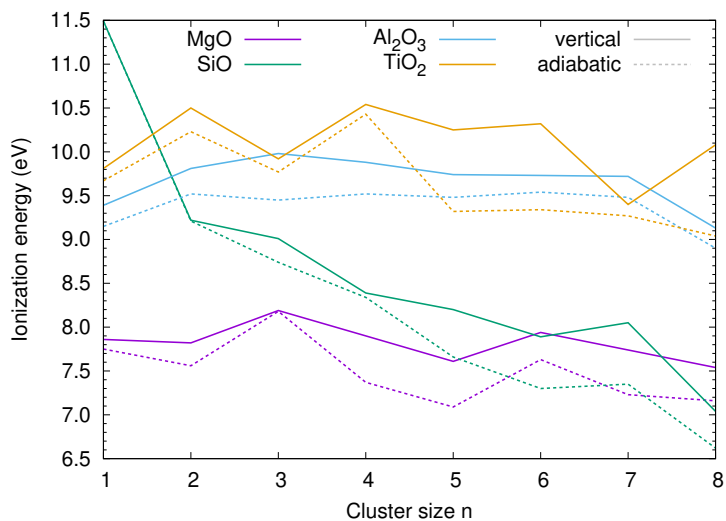}
\caption{The vertical (straight lines) and adiabatic (dashed lines) ionization energies (in eV) of the metal oxide clusters as a function of cluster size n \label{compa}.} 
\end{figure}

In Figure \ref{compa} the vertical and adiabatic ionization energies are represented graphically.
Magnesia clusters show a narrow range of 7.54$-$7.94 eV in their vertical ionization energies, apart from the trimer (n=3), which is explained in Section \ref{sadia}. These values are lower than the experimental value of 8.76$\pm$0.22 eV \citep{DALLESKA1994203} for the MgO\textsuperscript{+} cation. By inspecting the NIST Computational Chemistry Comparison and Benchmark Database (CCCBDB) database\footnote{http://cccbdb.nist.gov/} we find that the high level coupled cluster and the majority of the hybrid functionals underpredict the MgO ionization energies by about 1 eV. The deviation from the experimental result seems therefore to be inherent to DFT calculations. 
The largest ionization energy in our study (11.49 eV) is found for the SiO molecule, which is in excellent agreement with experimental results \citep{lemmon_etal:1998} giving a value of (11.49$\pm$0.02 eV).
Larger-sized (SiO)\textsubscript{n}, n$>$1, polymers show a decreasing trend with cluster size n, except for n=7.
Clusters of Al\textsubscript{2}O\textsubscript{3} show vertical ionization energies in a narrow range of 9.39$-$9.98 eV, apart for the largest considered size (n=8). To our knowledge there is no study that investigated  Al\textsubscript{2}O\textsubscript{3}\textsuperscript{+} experimentally. However, we note that the calculated alumina cluster ionization energies closely resemble the experimental value of 9.46$\pm$0.06 eV for the AlO molecule \citep{doi:10.1063/1.443274} and the value of 9.46 eV for its dimer, Al\textsubscript{2}O\textsubscript{2}, calculated by \citep{doi:10.1021/acs.jpca.0c07922}.
With the exception of the previously mentioned SiO molecule, the TiO\textsubscript{2} clusters exhibit the largest vertical ionization energies, ranging from 9.40 to 10.54 eV. As far as we know, there is no laboratory measurement of the TiO\textsubscript{2}\textsuperscript{+} cation and its related ionization energy.
However, the existence of the cation was proven experimentally \citep{https://doi.org/10.1002/anie.196610411}.
With the exception of (MgO)\textsubscript{5}, all four cluster families show their lowest adiabatic ionization energy at their largest cluster size (n=8). 

\subsection{Adiabatic ionization}
\label{sadia}
In the following, we present the structures of the adiabatically optimized cluster cations, compare them with the neutral clusters, and discuss their relative energies with respect to the vertical ionization energies. 
  
\subsubsection{Magnesia cations}
\begin{figure}[H]
\includegraphics[width=2.5 cm]{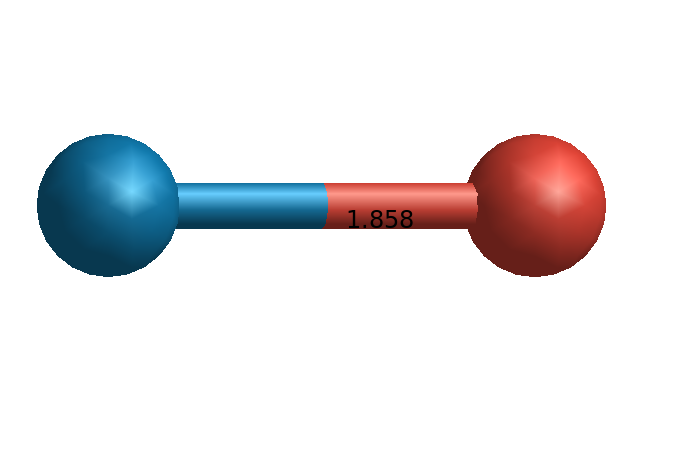}
\includegraphics[width=2.5 cm]{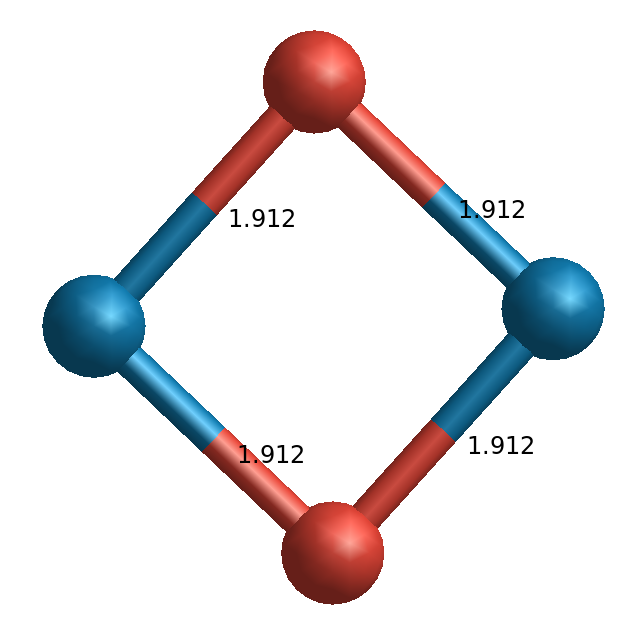}
\includegraphics[width=2.5 cm]{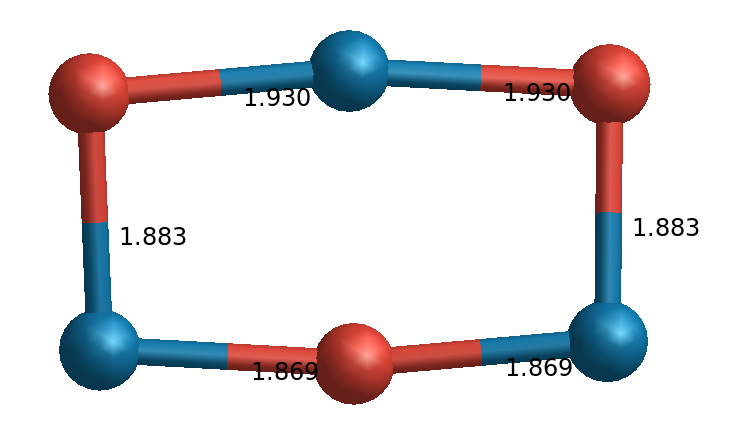}
\includegraphics[width=2.5 cm]{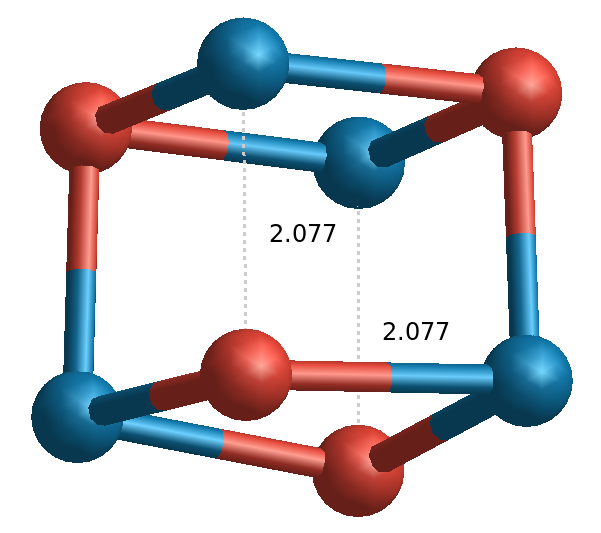}
\includegraphics[width=3.5 cm]{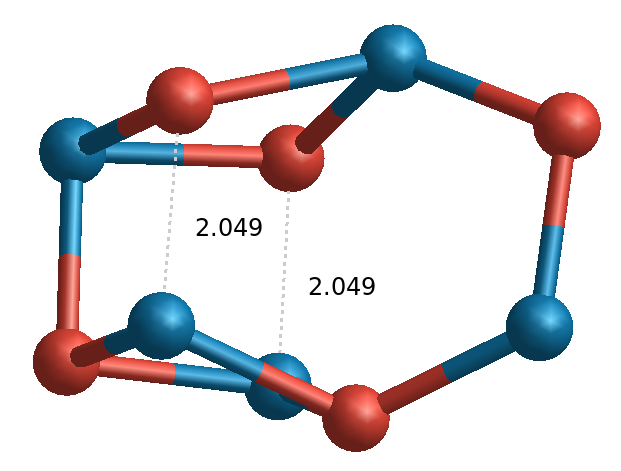}
\includegraphics[width=2.5 cm]{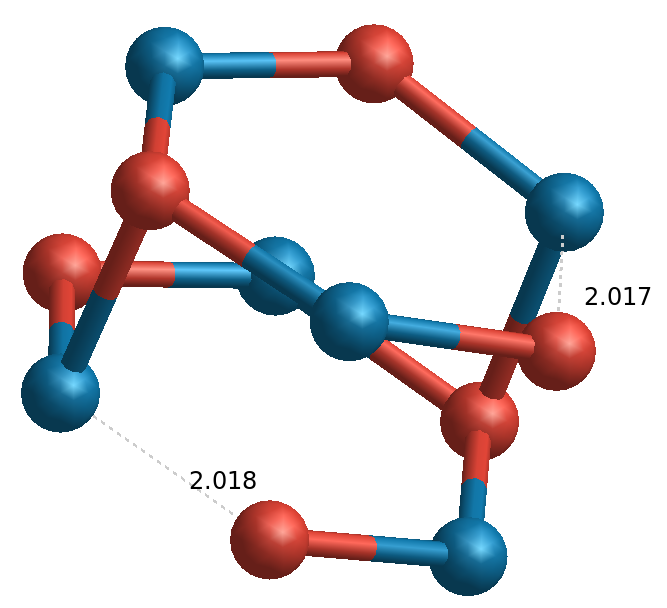}
\includegraphics[width=3.5 cm]{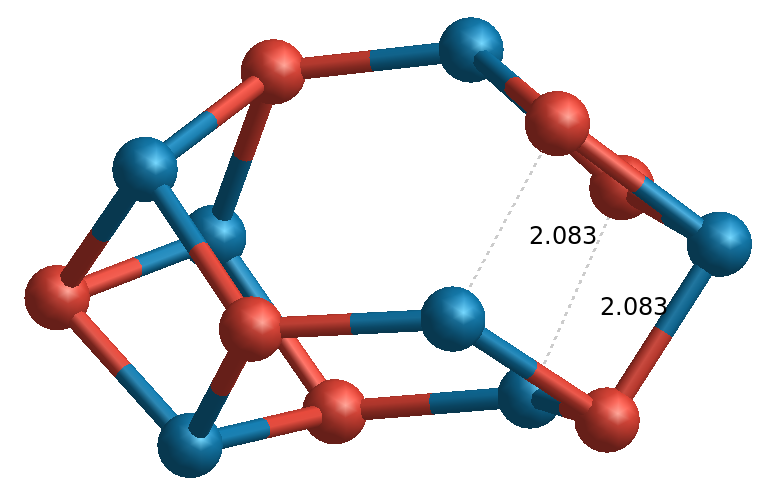}
\includegraphics[width=3.5 cm]{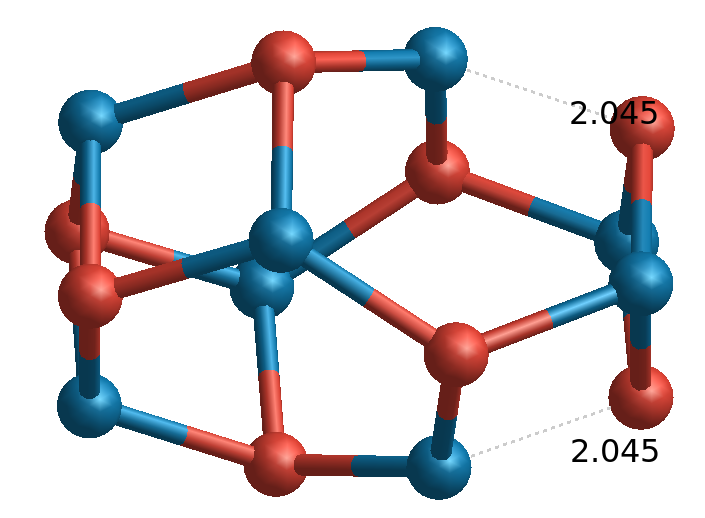}
\caption{The adiabatically optimized stuctures of (MgO)\textsubscript{n}\textsuperscript{+} cations. Mg atoms are in blue, O atoms in red, numbers correspond to bond distances in \AA{} \label{mgo+}.} 
\end{figure}

The adiabatically optimized MgO\textsuperscript{+} cation has a larger bond distance (d=1.858 \AA{}) than the neutral MgO molecule (1.767 \AA{}) as can be seen in Figure \ref{mgo+}.
The corresponding electron removal accounts for a difference of 0.09 \AA{} and 0.11 eV. 
Both, neutral (MgO)\textsubscript{2} and its cation, show flat square structures with equidistant edges. The cation has slightly larger bond lengths (1.912 \AA{}) as compared to the neutral dimer (1.892 \AA{}). It is thus impracticable to localize the electron removal through the ionization process.    
The neutral (MgO)\textsubscript{3} cluster structure is a ring consisting of 6 equidistant Mg$-$O bonds with a bond length of  1.852 \AA{}. Our initial searches for (MgO)\textsubscript{3}\textsuperscript{+} resulted in convergence failures. However, we found a rectangular-shaped cyclic structure to be a suitable candidate for the cationic form of (MgO)\textsubscript{3}. 
We note that a completely flat 2D geometry exhibits an imaginary vibration mode and is therefore likely to be a TS. 
A slight distortion of the geometry out of the 2D plane leads to a true minimum with only real vibration modes. The adapted different shape of (MgO)\textsubscript{3}\textsuperscript{+} is arguably the reason for the highest adiabatic ionization energy among the (MgO)\textsubscript{n} clusters.    
Neutral (MgO)\textsubscript{4} has the shape of a cube with identical edge (bond) lengths of 1.972 \AA{}. In its ionized state, the (MgO)\textsubscript{4}\textsuperscript{+} bond lengths are altered in a non-uniform way, 
where diagonally opposed edges have the same lengths. 
The longest two distances (marked with a dashed line) have a length of 2.077 \AA{} which is likely to be the location of the electron removal. 
In neutral (MgO)\textsubscript{5} the bond lengths are not identical, but show a range from 1.82 \AA{} to 2.00 \AA{}.
After removing an electron the (MgO)\textsubscript{5}\textsuperscript{+} bond lengths change in the region of the two rhombic substructures. Most prominent are two distances exceeding 2 \AA{} (marked with a dashed line) indicating the location of the electron disposal. 
The geometry of neutral (MgO)\textsubscript{6} consists of two “honeycomb” hexagons stacked on top of each other. The bond lengths within the two hexagons are 1.925 \AA{}, and the hexagons are connected by relatively  large distances (2.014 \AA{}). The optimization of (MgO)\textsubscript{6}\textsuperscript{+} cation leads to a TS. 
By distorting the TS geometry we find a real (MgO)\textsubscript{6}\textsuperscript{+} minimum showing two large Mg$-$O bonds of 2.02 \AA{}. Moreover, the hexagons in the cation are not plane-parallel as in the neutral n=6 cluster. 
For n=7, the bond lengths of the neutral cluster range from 1.90 to 1.97 \AA{}. The largest effects on the cation geometry are found at two Mg$-$O bonds with large bond lengths of 2.08 \AA{} (marked with dashed lines). 
The neutral (MgO)\textsubscript{8} cluster has a cylindrical shape with a square base (edge length 1.959 \AA{}). The remaining bond lengths range from 1.94$-$1.97 \AA{}. However, the cationic form (MgO)\textsubscript{8}\textsuperscript{+} did not converge, unless we distorted its geometry. The distorted geometry resembles the shape of a ``keyhole’’, corresponding to the second-lowest energy neutral (MgO)\textsubscript{8} isomer.
Therefore, we use the latter ``keyhole'' structure as a reference.
Its adiabatic ionization impacts most strongly two Mg$-$O bonds that exceed a distance of 2 \AA{} (marked with dashed lines).\\
For n=2, the site of the ionization could not be attributed to a specific location.
(MgO)\textsubscript{3}\textsuperscript{+} respresents a special case, as its structure was re-adjusted and thus, its ionization energy represents an outlier.
For n=4$-$8, the adiabatic ionization increases two Mg$-$O bonds simultaneously. 
Therefore, the ionization locations of the (MgO)\textsubscript{n}, n=2$-$8, could not be uniquely 
determined. This might be the result of electron delocalization, but it could also arise from approximating the electron probability by its density.


\subsubsection{Silicon monoxide cations}
\begin{figure}[H]
\includegraphics[width=2.5 cm]{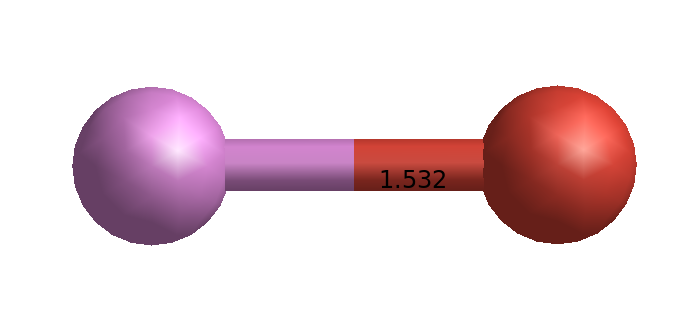}
\includegraphics[width=2.5 cm]{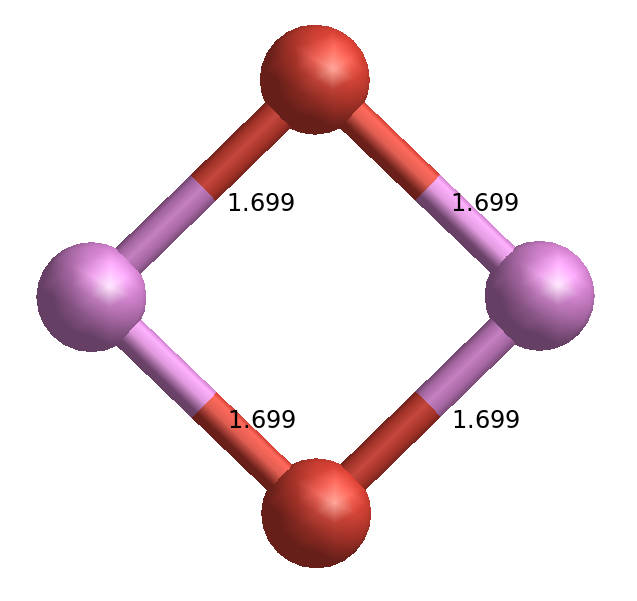}
\includegraphics[width=2.5 cm]{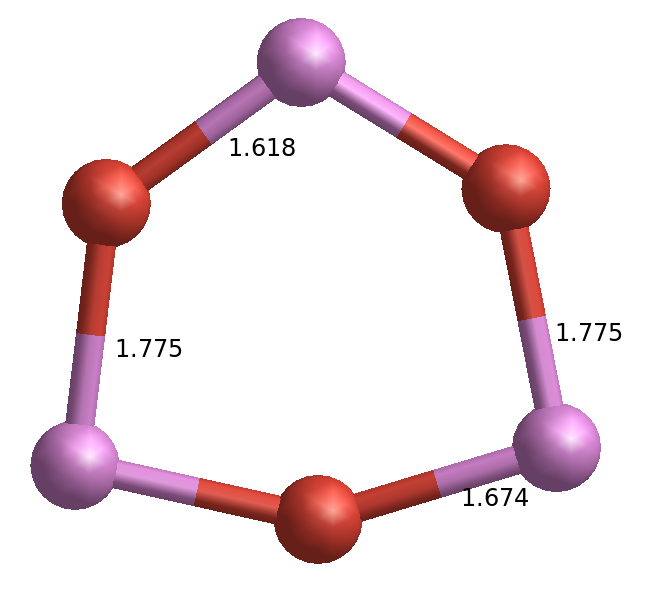}
\includegraphics[width=2.5 cm]{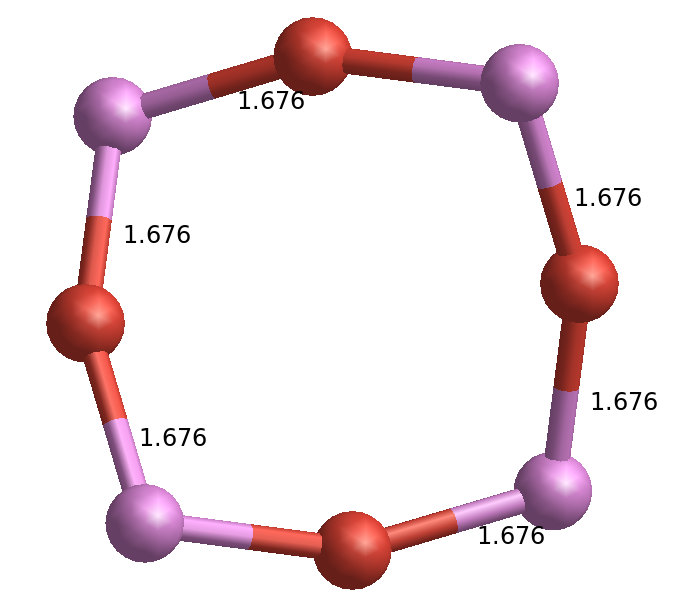}
\includegraphics[width=3.5 cm]{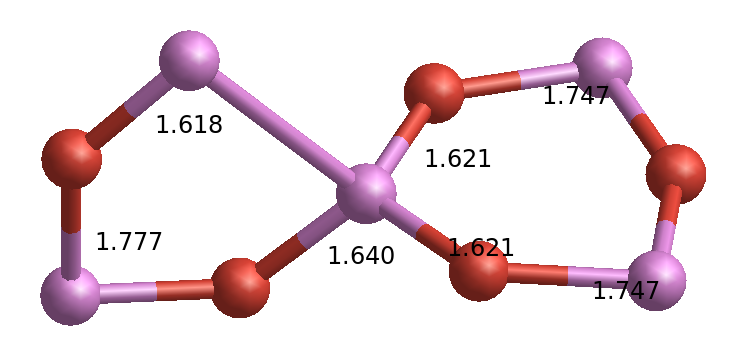}
\includegraphics[width=3.5 cm]{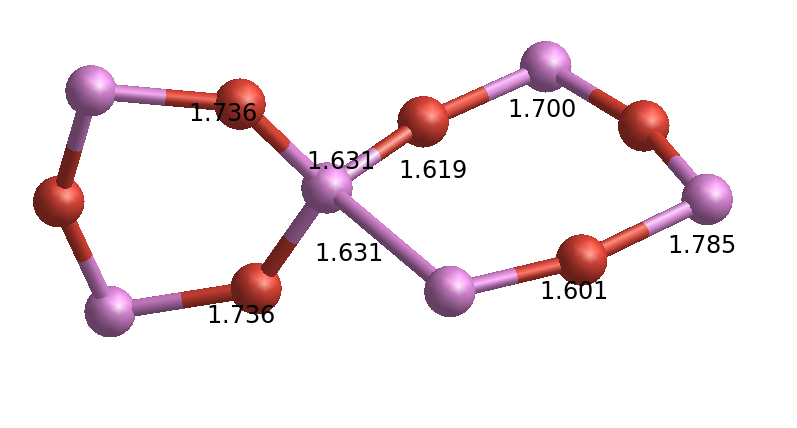}
\includegraphics[width=2.5 cm]{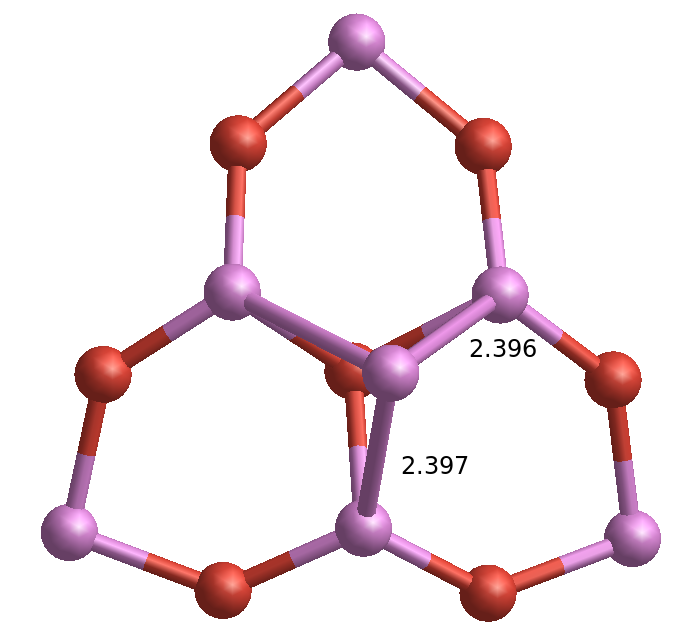}
\includegraphics[width=2.5 cm]{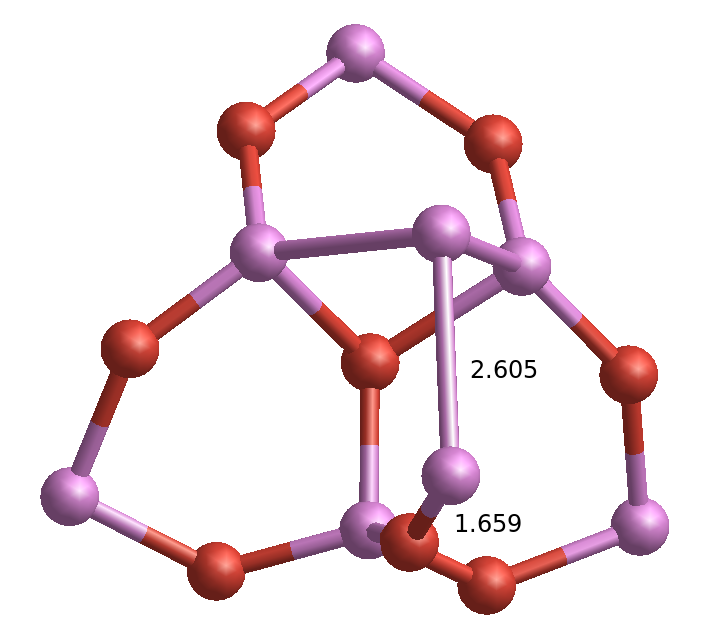}
\caption{The adiabatically optimized stuctures of 
(SiO)\textsubscript{n} cations. Si atoms are in violet, O atoms in red, numbers correspond to bond distances in \AA{} \label{sio}.}
\end{figure}

The Si$-$O bond length of the SiO\textsuperscript{+} cation is only marginally larger ($<$ 0.01 \AA{}) than that of the neutral SiO molecule (see Figure \ref{sio}). It is thus not surprising that vertical and adiabatic ionization energies have the same value (11.49 eV).
This ionization energy is the largest in our sample and reflects the particular stability of the SiO molecule.
The SiO dimer, (SiO)\textsubscript{2}, and its cation, (SiO)\textsubscript{2}\textsuperscript{+}, have both an ideal rhombic conformation with identical edge lengths. For the cation, they are slightly shorter (1.699 \AA{}) than for the neutral cluster (1.717 \AA{}). 
This difference accounts for just 0.01 eV of the large ionization energy of 9.21 eV.
In the (SiO)\textsubscript{3}\textsuperscript{+} cation two long (1.775 \AA{}) and two short (1.674 \AA{}) Si$-$O bonds appear, whereas they are all identical in the neutral (SiO)\textsubscript{3} cluster (1.691 \AA{}). This structural rearrangement lowers the energy by 0.27 eV.
Both, neutral and cationic (SiO)\textsubscript{4}\textsuperscript{(+)}, show identical lengths for all bonds. Their difference is very small (0.003 \AA{}). Also the bond angles of this non-planar geometry are hardly different ($<$4$^{\circ}$), which reflected in the small E\textsuperscript{v}-E\textsuperscript{a} difference of 0.05 eV.
The majority of the Si$-$O bonds in (SiO)\textsubscript{5}\textsuperscript{+} (i.e. where the lengths are indicated) display ionization-induced changes in their length by 0.05$-$0.07 \AA{} leading to a significant reduction of the energy (0.54 eV). The remaining three Si$-$O bonds are almost identical in the neutral and the cationic form. At this size (n=5) the adiabatic ionization energy has dropped below 8 eV and is thus similar to those of the (MgO)\textsubscript{n}.
As for n=5, there are three Si$-$O bonds that hardly change in (SiO)\textsubscript{6} through ionization. All other bonds (indicated by their bond lengths) deviate about 0.04$-$0.09 \AA{} as compared to the neutral cluster resulting in an energy decrease of 0.59 eV.  
For n=7, no large geometric changes are observed in the bond lengths (max. 0.03 \AA{} for Si$-$O bonds). However, the three appearing Si$-$Si bonds are shorter by 0.05 \AA{}, as compared with the neutral cluster, accounting for about 0.70 eV. This size represents an outlier in the trend of decreasing ionization energies with respect to the cluster size n.   

In (SiO)\textsubscript{8}\textsuperscript{+} only two bonds (one Si$-$Si, one Si$-$O, indicated by their bond lengths) have altered significantly in the ionization process accounting for 0.42 eV.\\
With the exception of n=7, (SiO)\textsubscript{n} clusters show decreasing vertical and adiabatic ionization energies with increasing cluster size n. Moreover, we find that for the small cation clusters (n=1$-$6), exhibiting strict cation-anion ordering, the ionization process could not (unambiguously) be attributed to a specific site. For the larger clusters (n=7$-$8) the electron removal could be localized towards the silicon segregation of the cluster. 
Our data set is too small to draw firm conclusions, but our study indicates that
the segregation of silicon impacts the ionization energies, possibly reducing them 
by supplying sites with smaller charge separation.

\subsubsection{Alumina cations}
\begin{figure}[H]
\includegraphics[width=3.0 cm]{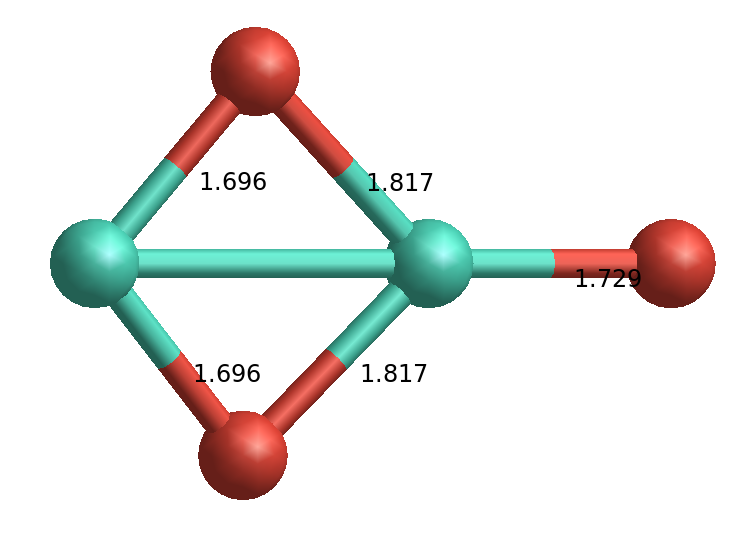}
\includegraphics[width=2.5 cm]{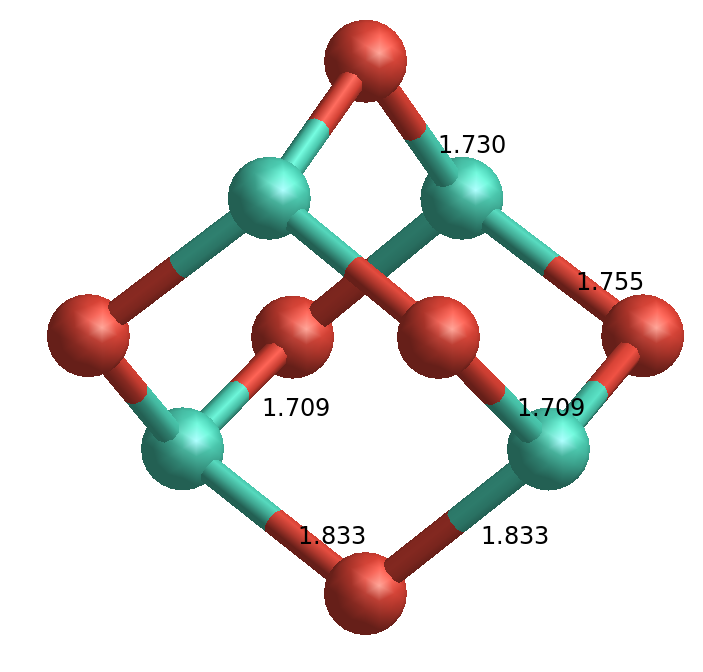}
\includegraphics[width=2.5 cm]{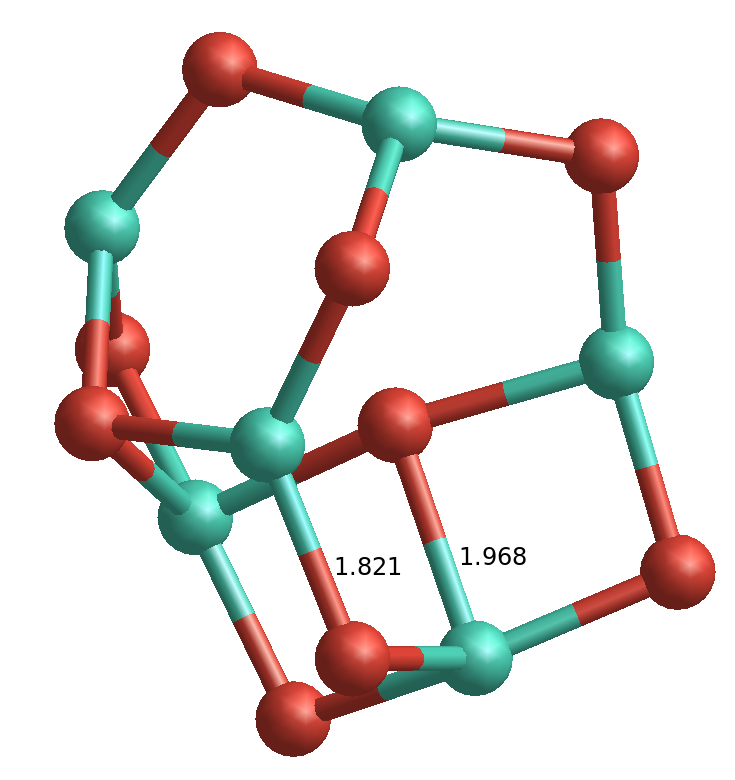}
\includegraphics[width=3.0 cm]{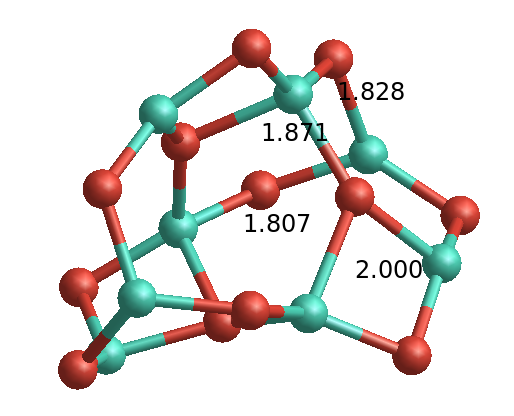}
\includegraphics[width=2.5 cm]{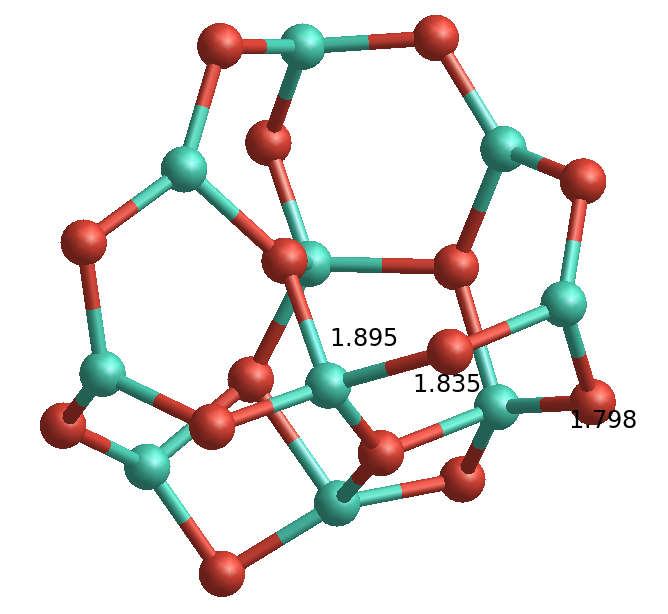}
\includegraphics[width=3.5 cm]{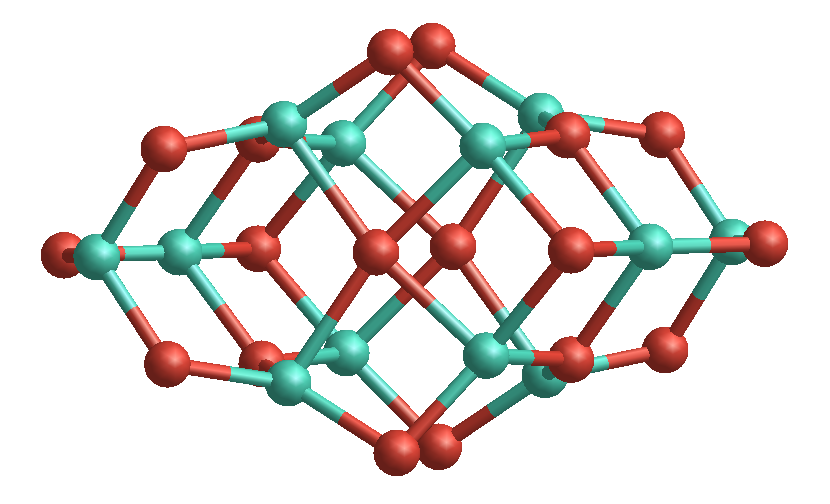}
\includegraphics[width=3.0 cm]{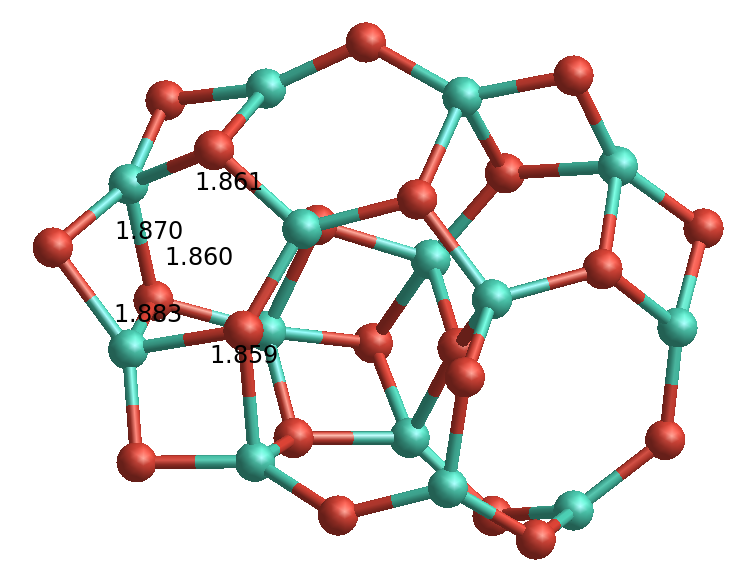}
\includegraphics[width=3.0 cm]{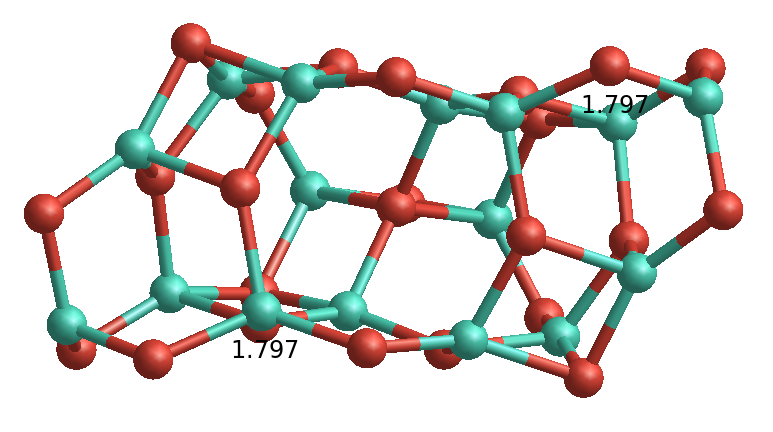}
\caption{The adiabatically optimized stuctures of 
(Al\textsubscript{2}O\textsubscript{3})\textsubscript{n}\textsuperscript{+} cations. Al atoms are in green, O atoms in red, numbers correspond to bond distances in \AA{} \label{fig4}.}
\end{figure}

The neutral Al\textsubscript{2}O\textsubscript{3} monomer is a triplet state with a ``kite''-shaped geometry. 
In contrast to the neutral cluster with quite similar Al$-$O bond lengths (1.747$-$1.777 \AA{}), the (Al\textsubscript{2}O\textsubscript{3})\textsuperscript{+} cation exhibits two longer bonds (1.817 \AA{}) and two shorter bonds (1.696 \AA{}) as can be seen in Figure \ref{fig4}. The removed electron originates thus potentially from the terminal Al atom, leaving another unpaired electron at the opposite site of the cluster (i.e. the tail of the kite). The monomer 
has an adiabatic ionization energy of 9.15 eV, which is lower than for larger alumina clusters sizes (with the exception of n=8).  
The adiabatic ionization of the highly symmetric neutral alumina dimer (point group T\textsubscript{d}) leads to a TS.
By distorting the TS geometry we find a true (Al\textsubscript{2}O\textsubscript{3})\textsubscript{2}\textsuperscript{+} minimum showing bonds with pairwise identical lengths and a corresponding lower symmetry (point group C\textsubscript{2}). The energy difference between the TS and the real cation minimum is just 0.03 eV.
For n=3, the largest change, induced by the adiabatic ionization, comes from a newly arising bond with a length of 1.968 \AA{} that is not present in the neutral cluster (distance 2.279 \AA{}). As a consequence of this new bond, the difference between vertical and adiabtic ionization (0.53 eV) is the largest among all considered alumina clusters.
The alumina tetramer cation shows four bonds that are enlarged by 0.07$-$0.10 \AA{} (indicated by their bond lengths) as compared to the neutral (Al\textsubscript{2}O\textsubscript{3})\textsubscript{4} cluster. 
The adiabatic and vertical ionization energies differ by 0.36 eV and account for 9.52 eV and 9.88 eV, respectively.  
In (Al\textsubscript{2}O\textsubscript{3})\textsubscript{5}\textsuperscript{+}, predominantly three Al$-$O bonds (lengths are given in Figure \ref{fig4}) are affected by the ionization process in changing their magnitude by 0.07$-$0.09 \AA{}.
The adiabatic ionization of (Al\textsubscript{2}O\textsubscript{3})\textsubscript{6}\textsuperscript{+} does not change the cluster geometry significantly, which is reflected by moderate adaptations in the bond lengths (max 0.04 \AA{}) and energies (0.19 eV). Moreover, the C\textsubscript{2h} symmetry of the cluster is preserved.
For n=7, the changes in the cluster geometry could be localized at one site (see bond length tags in Figure \ref{fig4}) and account for changes of 0.08$-$0.17 \AA{}.
The difference between adiabatic and vertical ionization energy is 0.24 eV.
The geometry of the (Al\textsubscript{2}O\textsubscript{3})\textsubscript{8}\textsuperscript{+} cation resembles closely to that of the neutral cluster. Two Al$-$O bonds deviate by 0.05 \AA{} (indicated by numbers). All other bonds change much less. The C\textsubscript{2} symmetry is preserved.\\ 
For alumina clusters of size n=2$-$7, the ionization energies are in a very narrow range of 9.45$-$9.54 eV, but are smaller for n=1 (9.15 eV) and n=8 (8.90 eV).
The comparatively low ionization potential of the monomer (n=1) can be explained by the disposal of an unpaired electron that does not constitute a bond. For n=8, we do not have a concise rationale, but it could be a size effect. All considered clusters families show lower energies for n=8 than for smaller sizes (n=1$-$7).

\subsubsection{Titania cations}
\begin{figure}[H]
\includegraphics[width=2.5 cm]{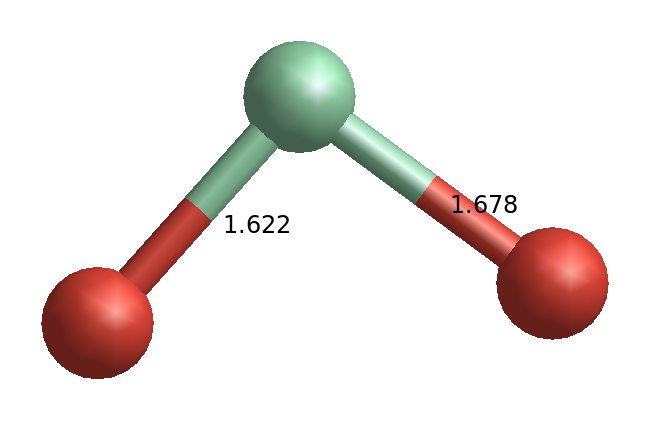}
\includegraphics[width=2.5 cm]{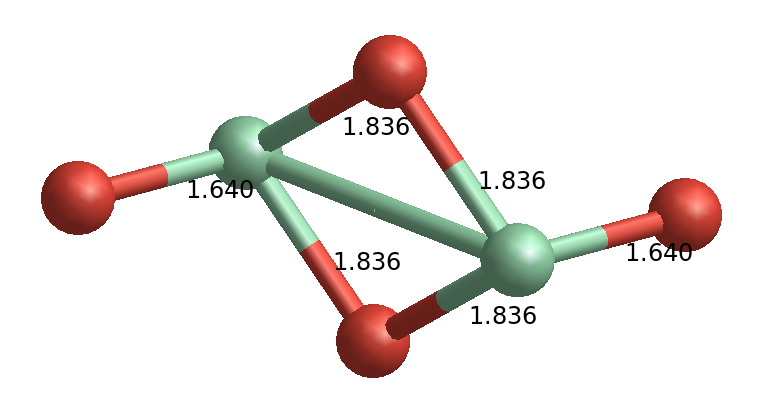}
\includegraphics[width=2.5 cm]{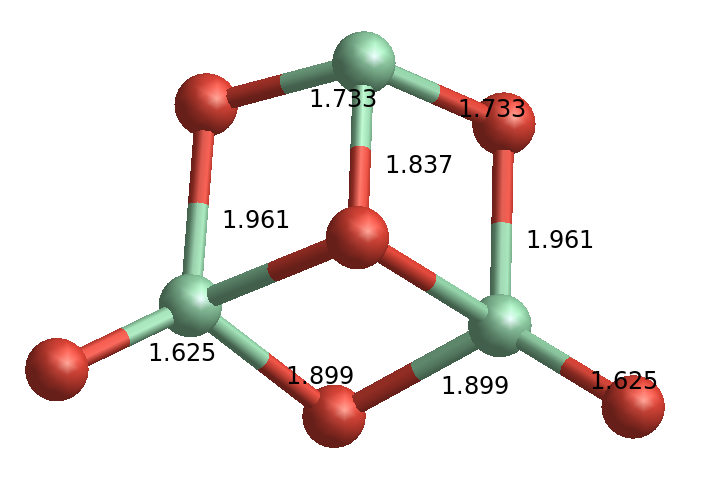}
\includegraphics[width=2.5 cm]{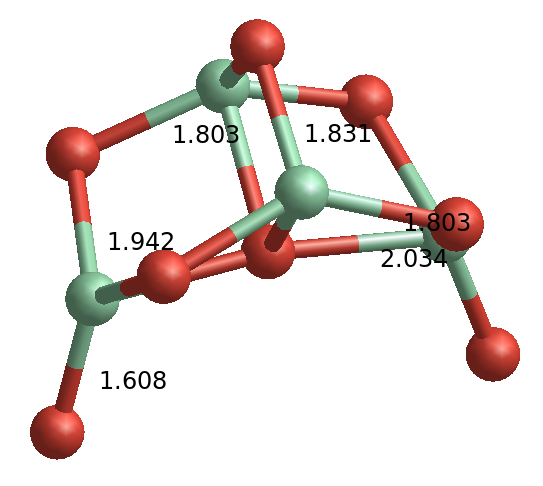}
\includegraphics[width=3.5 cm]{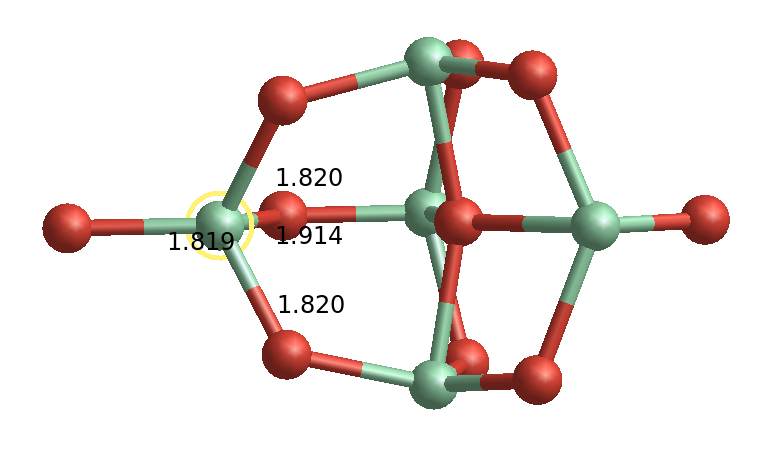}
\includegraphics[width=2.5 cm]{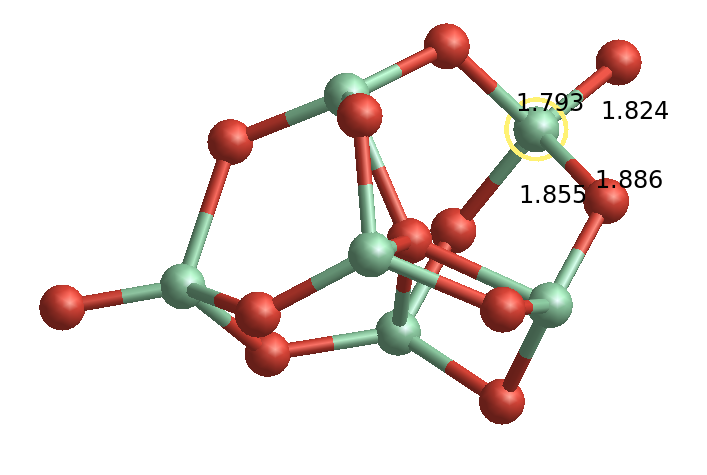}
\includegraphics[width=3.5 cm]{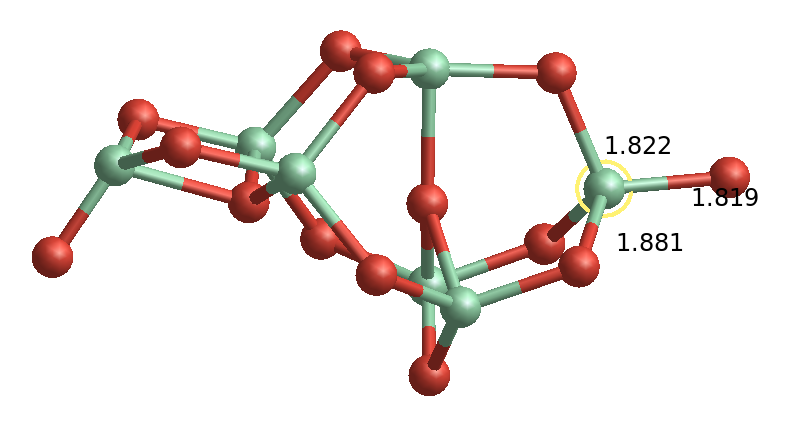}
\includegraphics[width=3.0 cm]{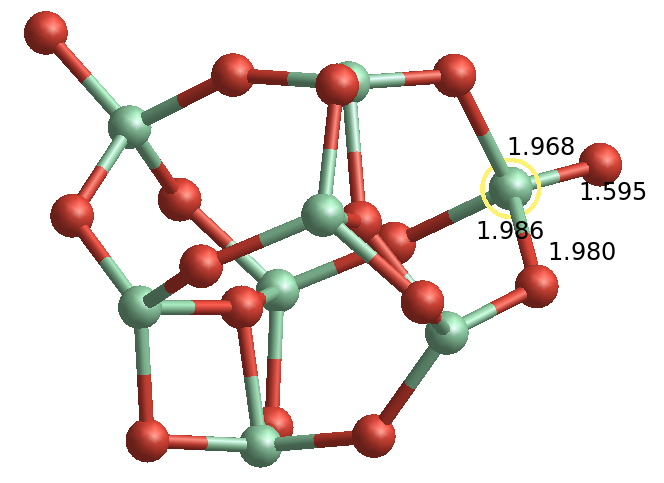}
\caption{The adiabatically optimized stuctures of 
(TiO\textsubscript{2})\textsubscript{n} cations. Ti atoms are in olive, O atoms in red, numbers correspond to bond distances in \AA{} \label{fig2}.}
\end{figure}

The adiabatic ionization of the TiO\textsubscript{2} molecule results in a TS exhibiting an imaginary vibrational frequency. 
By streching one of the two Ti$-$O bonds, we find a real TiO\textsubscript{2}\textsuperscript{+} minimum with two different bond lengths (1.678 \AA{} and 1.622 \AA{}, see Figure \ref{fig2}), as opposed to the neutral TiO\textsubscript{2} molecule, which shows two symmetric bonds with a length of 1.64 \AA{}. We note a smaller bond angle of 95.2$^{\circ}$ as compared to the neutral counterpart (111.7$^{\circ}$). The energy difference from the adiabatic optimization is 0.14 eV and the TS energy is less than 0.001 eV above the real minimum.
The titania dimer (n=2) hardly changes its geometry when it is adiabatically ionized and preserves its 
C\textsubscript{2h} symmetry. All six bonds change by less than 0.02 \AA{} accounting for 0.27 eV.
For (TiO\textsubscript{2})\textsubscript{3}, the overall geometry is not largely affected, since the cluster coordinates change by a maximum of 0.06 \AA{} through adiabatic ionization. 
The structure of the titania tetramer (n=4) cation is virtually not altered by the ionization process (max. 0.01 \AA{}), which is reflected in a moderate energy change (0.11 eV). 
The (TiO\textsubscript{2})\textsubscript{5}\textsuperscript{+} cation differs from the neutral (TiO\textsubscript{2})\textsubscript{5} mostly by one Ti atom (marked in yellow), whose four Ti$-$O bonds change by 0.08$-$0.21 \AA{} and are responsible for the energy difference of 0.93 eV.
In (TiO\textsubscript{2})\textsubscript{6}\textsuperscript{+}, one Ti atom (marked in yellow), which is located next to a terminal O atom, is most strongly affected by the adiabatic ionization accounting for almost 1 eV. This Ti atom shows different bond lengths ($\Delta$ d=0.09$-$0.21 \AA{}) than the neutral cluster.
The (TiO\textsubscript{2})\textsubscript{7}\textsuperscript{+} exhibits the most significant changes at one Ti atom (marked in yellow) that is connected to a terminal, single-bonded O atom. Here, the bond lengths change due to the ionization by 0.12$-$0.21 \AA{} and the energy by 0.13 eV. 
For n=8, the adiabatic ionization of the neutral GM candidate did not converge. Even the vertical ionization energy could not be determined. Therefore, we performed ionization calculations for a number of low-energy (TiO\textsubscript{2})\textsubscript{8} isomers. 
In the case of convergence, their adiabatic ionization energies are in the range of 8.76$-$9.21 eV. 
As a reference, we use the neutral (TiO\textsubscript{2})\textsubscript{8} isomer with the lowest relative energy (0.26 eV) with respect to the GM candidate. 
Its adiabatic ionization energy is 9.04 eV and is thus the lowest among the titania clusters. 
The largest geometric change is observed at one Ti atom, 
whose coordinates change by 0.11-0.22 \AA{} (bond lengths are indicated) and lower the energy considerably (by 1.04 eV).\\
The titania clusters show large adiabatic ionization energies ranging from 9.04$-$10.43 eV. Also the differences between vertical and adiabatic energies are non-negligible.
They exhibit an alternating pattern in their ionization energies.
Clusters with even size (n=2,4,6) show higher energies than their neighboring sizes with odd numbers (n=1,3,5,7).

\section{Discussion and Conclusion}
\label{discus}
Dust-forming metal oxide clusters show ionization energies that are lower than the one of atomic oxygen (13.6 eV),
but higher or comparable to the one of the corresponding metal.
MgO clusters show generally the lowest ionization energies which lie in a narrow range of 7.09$-$8.19 eV, overlapping with the  atomic Mg ionization energy of 7.65 eV.
Al\textsubscript{2}O\textsubscript{3} clusters exhibit ionization energies (8.9$-$10.0 eV) higher than that of atomic Al (5.99 eV), with the triplet monomer respresenting a special case.
TiO\textsubscript{2} clusters show the highest ionization energies (9.27$-$10.54 eV) among the considered metal oxides, exceeding the atomic Ti ionization potential of 6.83 eV. 
SiO clusters show the largest variation in their ionization energies (6.62$-$11.49 eV) that monotonically decrease with size, thereby crossing the atomic Si ionization energy of 8.15 eV. The decrease might be the result of the size-dependent atomic segregation of (SiO)\textsubscript{n} clusters.
All considered cluster families show a gradually declining trend of their ionization energies with cluster size. 
\\
Gas constituents usually have not a single temperature, but follow a statistical distribution function.\\

Assuming chemical equilibrium, we can assess the fraction of ionized clusters by using the Boltzmann law:

\begin{equation}
\frac{n(X^{+})}{n(X)}=\frac{2}{1} \exp(-\frac{E_{i}}{KT}) 
\label{saha}
\end{equation}

where $n(X^{+})$ and ${n(X)}$ are the number density of the ionized and neutral cluster species X, respectively. E$_{i}$ is the previously introduced ionization energy and $KT$ correspond to the thermal energy.
The factor $\frac{2}{1}$ arises from the spin multiplicity M of the neutrals (M=1) and cations (M=2) and accounts for the statistical weigths. Note that for the triplet X=Al\textsubscript{2}O\textsubscript{3}, it should read $\frac{2}{3}$.\\
From Table \ref{tab1} it is apparent that the ionization energies range from 6.62 eV to 11.49 eV corresponding to equivalent kinetic temperatures of 76 800$-$133 300 K. 
These temperatures are orders of magnitude larger than the prevailing photospheric temperatures of an AGB star, ranging typically between 2000 K and 3000 K. 
According to equation \ref{saha} the ionization fractions at the photosphere result in small values between 2.3$\times$10$^{-29}$ and 1.5$\times$ 10$^{-11}$ for T$\simeq$2000$-$3000 K.
However, pulsation-induced shocks, travelling periodically through the circumstellar envelopes of AGB stars, increase the temperature locally and temporarily. By applying the Rankine-Hugoniot jump conditions to photospheric conditions and a diatomic gas, temperatures $\sim$ 40000 K can be attained (see e.g. \citep{2016A&A...585A...6G}). This is still a factor of 2$-$3 smaller than the equivalent ionization temperatures. However, owing to its intrinsic chemical equilibrium assumption, equation \ref{saha} is not applicable to an immediate post-shock gas strongly deviating from equilibrium conditions.\\
Finally, we want to address radiation fields. Stellar radiation fields from AGB stars with temperatures of 2000$-$3000 K typically peak at micrometer wavelengths in the infrared. Though the majority of the stellar photons is thus not capable of ionizing the presented metal oxide clusters, there is a certain probability that UV photons originating from the high energy tail of their distribution induce ionization.
Although the ionization fraction is comparatively low at characteristic AGB temperatures, the orders-of-magnitude faster ion-molecule chemistry with rates up to 10$^{0}$ cm$^3$s$^{-1}$ could compete with neutral-neutral reaction rates that usually range between 10$^{-9}-$10$^{-13}$ cm$^3$s$^{-1}$. 
Accounting for both, the fast kinetic rates (i.e. 9$-$13 orders of magnitude) and the low abundances of the cluster cations (i.e. 11$-$29 orders of magnitude), we find that the ion-molecule chemistry can compete with neutral-neutral reactions for the lowest ionization energies (< 8 eV) and the fastest ion-molecule rates.  Moreover, pulsation-induced shocks are prone to increase the amount of ions and therefore also enhance their impact on the circumstellar chemistry. Therefore, an ion-molecule driven chemistry can become important in the dust formation zone of AGB stars, in particular in the immediate post-shock regime. 
However, neutral-neutral reactions are still believed to represent the  dominant nucleation mechanism, as only extreme cases (i.e. reactions with the fastest ion-molecule rates and the largest ion concentrations  corresponding to low ionization energies) occur on comparable timescales.

We conclude that albeit ionization of the metal oxide clusters is of secondary importance under equilibrium conditions, non-equilibrium effects like periodic shocks, or radiation from a hot companion star
can provide the energy required to ionize (part of) the metal oxide clusters. Moreover, the (inter-)stellar radiation field is a promising source of photons with sufficiently large energies capable of ionizing metal oxide clusters. \citep{Van_de_Sande_2019} investigated the effect of stellar UV photons on the chemistry occuring in AGB winds and found that photo-dissociation can change considerably the related molecular abundances. It is thus throughout viable, that also photo-ionization can take place.
The presence of metal oxide cations would imply significant effects on the  nucleation pathways, its related energies and growth rates.
A realistic description of the temperature structure and radiation field in the highly dynamic envelopes of AGB stars is, however, beyond the scope of this paper. \\

\end{paracol}
%
%
%

%




\authorcontributions{All authors have read and agreed to the published version of the manuscript.}

\funding{D.G. and L.D. acknowledge funding by the ERC consolidator grant number 646758. J.-P. S. and H. L.-M. acknowledge the European Union's Horizon 2020 research and innovation programme under the Marie Sklodowska-Curie grant agreement No 860470.}




\dataavailability{Data is contained within the article or supplementary material
The data presented in this study are available at http://dave202.bplaced.net/ions/} 

\acknowledgments{We acknowledge the CINECA award under the ISCRA initiative, for the availability of high performance computing resources and support}

\conflictsofinterest{The authors declare no conflict of interest.} 

\reftitle{References}


\externalbibliography{yes}
\bibliography{biblio}

\end{document}